\documentclass[letterpaper, 10 pt, conference]{ieeeconf}  

\IEEEoverridecommandlockouts                              
\usepackage{textcomp}
\usepackage{amsthm}
\usepackage{amsmath,amssymb,amsfonts}
\allowdisplaybreaks
\usepackage{graphicx}  
\usepackage{cite}
\usepackage{nicefrac}
\usepackage[english]{babel}
\usepackage{cases}
\usepackage{algorithm, algorithmic}
\usepackage[caption=false,font=footnotesize]{subfig}
\usepackage{hyperref}
\hypersetup{
    colorlinks = true,
    citecolor = {blue},
    linkcolor=blue
}
\usepackage{dsfont}
\newcommand{\one}{\mathds{1}}

\newtheorem{theorem}{Theorem}

\newtheorem*{assumption*}{\assumptionnumber}
\providecommand{\assumptionnumber}{}
\makeatletter
\newenvironment{assumption}[1]
{\renewcommand{\assumptionnumber}{A#1}\begin{assumption*}\protected@edef\@currentlabel{A#1}}
	{\end{assumption*}
}
\makeatother

\newcommand{\tr}{^\mathrm{T}}

\newcommand{\norm}[1]{\|#1\|}
\newcommand{\indnorm}[1]{{\left\vert\kern-0.25ex\left\vert\kern-0.25ex\left\vert #1 
		\right\vert\kern-0.25ex\right\vert\kern-0.25ex\right\vert}}
\newcommand{\defeq}{\doteq}
\newcommand{\confreg}[1][2]{\mathcal{C}_{#1}}

\newcommand{\BR}{\mathbb{R}}
\newcommand{\BE}{\mathbb{E}}
\newcommand{\BP}{\mathbb{P}}
\newcommand{\CA}{\mathcal{A}}
\newcommand{\CB}{\mathcal{B}}
\newcommand{\CG}{\mathcal{G}}

\newcommand{\CN}{\mathcal{N}}
\newcommand{\CC}{\mathcal{C}}
\newcommand{\CE}{\mathcal{E}}

\DeclareMathOperator*{\argmin}{argmin}
\usepackage{xcolor}

\makeatletter
\newcommand\fs@spaceruled{\def\@fs@cfont{\bfseries}\let\@fs@capt\floatc@ruled
  \def\@fs@pre{\vspace{0.5\baselineskip}\hrule height.8pt depth0pt \kern3pt}\def\@fs@post{\kern3pt\hrule\vspace{-2mm}\relax}\def\@fs@mid{\kern3pt\hrule\kern2pt}\let\@fs@iftopcapt\iftrue}
\makeatother

\def \newtext{}

\title{\LARGE \bf
Non-Asymptotic Confidence Regions for Separable Nonlinear Models
}

\author{Szabolcs Szentpéteri \and Péter Kovács, \IEEEmembership{Senior Member, IEEE} \and Balázs Csan\'ad Csáji, \IEEEmembership{Senior Member, IEEE}\thanks{Sz.~Szentpéteri and B.~Cs.~Csáji were supported by the ``Robust Uncertainty Quantification for Learning and Control'' ADVANCED project of the National Research, Development and Innovation Office of Hungary (NKFIH), grant number 153390.
P.~Kov\'acs was supported by the J\'anos Bolyai Research Scholarship of the Hungarian Academy of Sciences (MTA) and the University Excellence Fund of E\"otv\"os Lor\'and University (ELTE).}\thanks{Sz.~Szentpéteri is with the Department of Automatic Control, Lund University, Sweden. He is a member of the ELLIIT Strategic Research Area at Lund University. This work was carried out while he was with the Institute for Computer Science and Control (SZTAKI), Hungarian Research Network (HUN-REN), 13--17 Kende utca, 1111, Budapest, Hungary {\tt\small szabolcs.szentpeteri@control.lth.se} }\thanks{P.~Kov\'acs is with the Department of Numerical Analysis, E\"otv\"os Lor\'and University (ELTE),
1117 Budapest, Hungary {\tt\small kovika@inf.elte.hu}}\thanks{B.~Cs.~Csáji is with the Institute for Computer Science and Control (SZTAKI),
		Hungarian Research Network (HUN-REN), 13--17 Kende u., 1111, Budapest, Hungary, and with the Department of Probability Theory and Statistics, Institute of Mathematics, E\"otv\"os Lor\'and University (ELTE),
		1/C Pázmány Péter sétány, 
        1117, Budapest, Hungary {\tt\small csaji@sztaki.hu}}}

\begin{document}

\maketitle
\thispagestyle{empty}
\pagestyle{empty}

\begin{abstract}
This paper studies robust uncertainty quantification for separable nonlinear least squares (SNLLS) problems.
We propose two identification methods, (i) Residual-Perturbed Gradient (RPG) and (ii) Global-Fit-Gap (GFG), to construct joint distribution-free confidence regions for both the linear and nonlinear parameters.
{\newtext By exploiting an assumed group invariance of the noise (such as symmetry or exchangeability), we apply random perturbations (such as sign-changes or permutations) to build the regions based on resampling and ranking.}
{\newtext We prove that both methods guarantee non-asymptotically exact coverage (type I error).}
{\newtext Finally, we illustrate the constructions through two numerical experiments: (i) identifying generalized finite impulse response (GFIR) systems using Laguerre filters with an adjustable decay rate and (ii) fitting Gaussian radial basis function (RBF) models with variable knots and widths.}
\end{abstract}

\section{Introduction}

Many {\em inverse problems} arising in machine learning, system identification, and signal and image processing can be formulated as {\em separable nonlinear least-squares} (SNLLS) problems \cite{SNLLapp1}.
In these models, the parameters can be naturally partitioned into a {\em nonlinear} component and a {\em linear} component. Typically, the nonlinear parameters form a relatively small set with direct physical interpretation, such as absorption coefficients, dominant frequencies, or phases, while the linear parameters correspond to coefficients associated with basis functions or system responses.

A widely used approach for solving SNLLS problems is the {\em variable projection} (VP) method~\cite{Golub1973}, which reduces the dimensionality of the search space and transforms the original problem into a lower-dimensional nonlinear optimization problem that is often easier to solve and numerically more stable. VP methods are particularly attractive for a wide range of inverse problems~\cite{golub2003}, including time-series prediction using RBF-ARX models~\cite{VP_RBF-ARX}, frequency identification~\cite{VPident}, blind image deconvolution, and image denoising~\cite{SNLLapp1}. Non-smooth~\cite{VP_L1, nonsmoothVP} and large-scale~\cite{VPlargescale} extensions have only recently been considered, further highlighting the potential of the SNLLS framework and the VP approach.

Golub and Pereyra showed that the {\em gradient} of the VP objective functional can be computed explicitly~\cite{Golub1973}, which has attracted considerable attention from the machine learning community. In particular, the VP algorithm can be used to train feedforward neural networks~\cite{VP_machine_learning}. {\newtext Recently, this approach has been generalized
to non-quadratic objectives,
allowing VP-based training of neural networks for tasks including segmentation and classification~\cite{VPtrain}. The VP operator has also been used as an activation mechanism that integrates domain knowledge into 
network architectures through a layer-wise SNLLS formulation of the target problem~\cite{WVPNet}.}

{\newtext VP methods are very efficient to compute {\em point estimates},  selecting a single model from a given class.
However, for {\em risk-aware} solutions, these should be augmented with {\em confidence regions} that quantify the {\em uncertainty} of likely models.}

Despite its {\em critical importance}, there are very few methods that address the issue of constructing confidence regions around VP estimates. Currently, the prevailing general solution relies on the {\em asymptotic distribution} of the scaled estimation error \cite{kundu1999asymptotic,bertozzi2024asymptotic}.
However, ensuring the existence of the limiting distribution is nontrivial for complex nonlinear structures \cite{vdVaart1998}. More importantly, the resulting confidence ellipsoids are only asymptotically guaranteed, providing a false sense of security in the finite-sample regime.

The {\em main contributions} of this paper are as follows.
\begin{itemize}
\item[(1)] 
{\newtext We introduce two methods for building {\em distribution-free} confidence sets for nonlinear models, centered around VP estimates. The {\em Residual-Perturbed Gradient} (RPG) method is computationally efficient, but it might capture artifacts.
The {\em Global-Fit-Gap} (GFG) method mitigates this issue, but incurs a higher computational cost.}
\item[(2)]
{\newtext We prove that both methods ensure confidence sets with {\em non-asymptotic} coverage guarantees, under only mild assumptions on the noises affecting the system.}
\item[(3)] Finally, we present two case studies: (i) identifying GFIR models for systems characterized by slowly decaying impulse responses, and (ii) estimating Gaussian-based RBF models with adjustable knots and widths.
\end{itemize}

\section{Problem Setting}
{\newtext In this section, we present the addressed SNLLS problem.}
\subsection{Data Generation \newtext and Problem Statement}
{\newtext Consider the following {\em separable nonlinear model}:
\begin{equation}
    Y_t \defeq \sum_{i=1}^{d_{\gamma}} \gamma^*_i \varphi_i(\beta^*,X_t) + W_t,
\end{equation}
where
{\newtext $\gamma^*\defeq[\gamma_1^*,\dots, \gamma_{d_{\gamma}}^*]\tr\in\BR^{d_{\gamma}}$} is the true {\em linear} parameter, $\beta^* \in \BR^{d_{\beta}}$ is the true {\em nonlinear} parameter, $\varphi_i(\beta^*,X_t)$ is a nonlinear basis function, $X_t \in \BR^{d_x}$ is the input,
$W_t\in \BR$ is the random noise, and $Y_t\in \BR$ is the output. We are given a sample of size $n$ of inputs and outputs $\{(X_t,Y_t)\}_{t=1}^n$.

We use the stacked notation $\theta \defeq [\beta\tr, \gamma\tr]\tr \in \BR^{d_{\theta}}$, where $d_{\theta} = d_{\gamma}+d_{\beta}$.
This convention applies equally to parameters, estimators, etc.
{\newtext For $k \in \mathbb{N}$, we also write $[k] \defeq \{1,\dots,k\}$.}

Using the input-output data, we address the problem of constructing a confidence region $\CE_p$ for the true parameter $\theta^*$ such that, for a prescribed confidence probability $p\in(0,1)$,
\begin{equation}
    \BP(\theta^* \in \CE_p) \geq p.
\end{equation}

We consider {\em deterministic} inputs; however, our results can be straightforwardly generalized to {\em exogenous} inputs, where the input and the noise sequences are independent.}

\subsection{Assumptions}
Our {\em main assumptions} on the noises affecting the system and on the nonlinear basis functions are as follows.
\begin{assumption}{1}\label{assu:noise}
{\newtext The noise $w \defeq [W_1,\dots, W_n]\tr$ is distributionally
invariant w.r.t.\ a known compact matrix group of transformations, $(\CG, \circ)$, where each $G \in \CG$ acts on $\BR^n$ and the composition $\circ$ is the group operation, i.e., $\forall\, G \in \CG: G(w) \stackrel{d}{=} w$.}
\end{assumption}

This assumption states that there exist known transformations that leave the {\em joint distribution} of the noise vector unchanged. Standard examples for which we know such transformations include {\em exchangeable} sequences, which in particular cover the {\em independent and identically distributed} (i.i.d.) case, and independent sequences that are {\em symmetric} about zero. 
{\newtext In the exchangeable case, the relevant transformations are multiplications by {\em permutation} matrices:
\begin{equation}\label{equ:G_i_permutation}
    \mathcal{G} = \{\hspace{0.3mm} G \in \{0,1\}^{n\times n} : G\one=\one,\; G\tr \one=\one \hspace{0.3mm}\},
\end{equation}
where
$\one$ is the all-one vector. In the independent and symmetric case, the transformations correspond to
multiplications by diagonal ``{\em sign}'' matrices, i.e., with $\pm 1$ entries:
\begin{equation}\label{equ:G_i_sign_perturbation}
    \hspace*{-2mm}\mathcal{G} = \{\hspace{0.3mm} \textrm{diag}(\alpha_{1},\dots,\alpha_{n}) :
    \forall k\in[n], \alpha_k \in \{-1, 1\}
    \hspace{0.3mm}\},
\end{equation}
where $\textrm{diag}(v)$ denotes a diagonal matrix with 
$v$ on its main diagonal.
Both groups are finite, thus compact.
{\newtext In both cases, the group operation is matrix multiplication,
the identity 
is the identity matrix, and the inverse is the matrix inverse.}

It is important to note that Assumption~\ref{assu:noise} is very mild: it imposes no conditions either on the {\em tail behavior} of the noise distribution or on the existence of its {\em moments}.}

\begin{assumption}{2}\label{assu:full_rank_Jacobian}

Let $\Upsilon \subseteq \mathbb{R}^{d_\beta}$ be open. 
{\newtext For $i\in [d_\gamma]$,} the nonlinear basis functions
$\varphi_i : \Upsilon \times \mathbb{R}^{d_x} \to \mathbb{R}$
are continuously differentiable 
w.r.t.\ $\beta$.
Define the regression matrix $\Phi(\beta) \in \mathbb{R}^{n \times d_\gamma}$ for 
$t\in[n]$ and $i\in[d_\gamma]$ as $\Phi(\beta)_{t,i} \defeq \varphi_i(\beta,X_t)$.
Let $J_{\beta'}(\Phi(\beta')\gamma'):\Upsilon \times \mathbb{R}^{d_\gamma}\to \mathbb{R}^{n\times d_\beta}$ be the Jacobian of $\Phi(\beta')\gamma'$ w.r.t.\ $\beta'$. We assume that for all $\beta \in \Upsilon$, $\gamma \in \mathbb{R}^{d_\gamma}$, and $n \ge d_\gamma + d_\beta$:
{\newtext $\mathrm{rank}
([J_{\beta'}(\Phi(\beta')\gamma')(\beta, \gamma), \Phi(\beta)])
=
d_\gamma + d_\beta$.}
\end{assumption}

This assumption ensures that the model is locally {\em identifiable} and
that the least-squares problem, detailed in the next {\newtext section}, is {\em well-posed}. It also guarantees that the confidence regions generated by our two proposed methods are {\em bounded} under suitable assumptions on the random transformations.
\section{Separable Nonlinear Least Squares}
\subsection{Least-Squares Estimate}
Given the sample $\{(X_t,Y_t)\}_{t=1}^n$, the parameters $\gamma^*, \beta^*$ can be estimated by minimizing the {\em least-squares} (LS) functional
\vspace*{-2mm}
\begin{equation}\label{equ:snlls_scalar_form}
    \argmin_{\beta,\gamma}\frac{1}{2}\sum_{t=1}^n\!\bigg(Y_t-\sum_{i=1}^{d_{\gamma}} \gamma_i \varphi_i(\beta,X_t)\bigg)^{\!2}\!.
\end{equation}
This problem is known as the {\em separable nonlinear least-squares} (SNLLS) problem.
By defining the residual as
\begin{equation}
   r(\theta) \defeq r(\beta,\gamma) \defeq y-\Phi(\beta)\gamma,
\end{equation}
where 
$y \defeq [Y_1,\dots, Y_n]\tr$, the LS problem of \eqref{equ:snlls_scalar_form} can be written in a compact matrix form as
\begin{equation}\label{equ:snlls_matrix_form}
    \hat{\theta}\defeq\argmin_{\theta \in \BR^{d_{\theta}} }\frac{1}{2}\|r(\theta)\|^2 =\argmin_{\beta,\gamma}\frac{1}{2}\|r(\beta,\gamma)\|^2.
\end{equation}
\subsection{Variable Projection}
A well-known approach to solve the SNLLS problem is the {\em variable projection} (VP) algorithm~\cite{Golub1973, golub2003}. This method first solves the linear LS problem for a fixed $\beta$ as
\begin{equation}
    \hat{\gamma}(\beta) = \argmin_{\gamma}\frac{1}{2}\|r(\beta,\gamma)\|^2 = \Phi(\beta)^\dagger y,
\end{equation}
where $(\cdot)^\dagger$ denotes pseudoinverse (Moore--Penrose inverse).
Then, it minimizes the following VP functional
\begin{equation}\label{equ:reduced_functional}
    \argmin_{\beta}\frac{1}{2}\|r_2(\beta)\|^2 = \argmin_{\beta}\frac{1}{2}\|(I-\Phi(\beta)\Phi(\beta)^\dagger)y\|^2,
\end{equation}
which can be obtained by substituting $\hat{\gamma}(\beta)$ to the original functional in \eqref{equ:snlls_matrix_form}.
Golub and Pereyra \cite{Golub1973} showed that the analytical expression for the $k$-th column of the {\em Jacobian} of the reduced functional $J_{\beta'}(r_2(\beta'))(\beta)$
is given as
\begin{align}
J_{\beta'}^{(k)}(r_2(\beta'))(\beta)&=-\Big[P_{\Phi(\beta)}^\perp \frac{\partial \Phi(\beta)}{\partial \beta_k}\Phi(\beta)^\dagger\notag\\
&\quad+\big(P_{\Phi(\beta)}^\perp \frac{\partial \Phi(\beta)}{\partial \beta_k} \Phi(\beta)^\dagger\big)\tr\Big]y,
\end{align}
where $P_{\Phi(\beta)}^{\perp} \defeq I-\Phi(\beta)\Phi(\beta)^\dagger$.
The Jacobian matrix of the reduced residual $J_{\beta'}(r_2(\beta'))(\beta)$ plays a key role in computing the gradient and the Hessian of the VP functional, solving the optimization problem of \eqref{equ:reduced_functional} and obtaining estimate $\hat{\beta}$.
\subsection{Gradients of the Residual}
In \cite{Golub1973}, the {\em gradients} of $\frac{1}{2}\|r(\theta)\|^2$ are also given as:
{\newtext
\begin{align}\label{equ:derivs_res_norm}
    \nabla_{\beta} \frac{1}{2}\|r(\theta)\|^2 &= \nabla_{\beta} \frac{1}{2}\|r(\beta,\gamma)\|^2 \notag\\
    &= -J\tr_{\beta'}(\Phi(\beta')\gamma')(\beta, \gamma) r(\beta,\gamma),\\
    \nabla_{\gamma} \frac{1}{2}\|r(\theta)\|^2 &= \nabla_{\gamma} \frac{1}{2}\|r(\beta,\gamma)\|^2 = -\Phi(\beta)\tr r(\beta,\gamma),
\end{align}}
therefore, we have
{\newtext
\begin{align}
    \nabla_{\beta} \frac{1}{2}\|r(\beta,\gamma)\|^2 \Big|_{(\beta^*, \gamma^*)} &= -J\tr_{\beta'}(\Phi(\beta')\gamma')(\beta^*, \gamma^*) w, \\
    \nabla_{\gamma} \frac{1}{2}\|r(\beta,\gamma)\|^2 \Big|_{(\beta^*, \gamma^*)} &= -\Phi(\beta^*)\tr w,\label{equ:res_derivs_at_true}
\end{align}}
and the Jacobian of $r(\theta) = r(\beta,\gamma)$ is
{\newtext
\begin{align}\label{equ:jacobian}
    J(\theta)&\defeq \left[ J_{\beta'}(r(\beta',\gamma')) (\beta, \gamma), \, J_{\gamma'}(r(\beta',\gamma')) (\beta, \gamma)\right]\notag\\
    &=-\begin{bmatrix} J_{\beta'}(\Phi(\beta')\gamma')(\beta, \gamma), \, \Phi(\beta)\end{bmatrix}.
\end{align}}
We also introduce the notation
\begin{align}\label{equ:def_Rn}
    R_n(\theta) &\defeq R_n(\beta, \gamma) = \frac{1}{n}J(\theta)\tr J(\theta),
\end{align}
which matrix plays an important role in the asymptotic theory of 
nonlinear least-squares problems presented later.
Note that under Assumption~\ref{assu:full_rank_Jacobian}, $\det(R_n(\theta))\neq 0$ for every $n \geq d_{\theta}$.

\subsection{Asymptotic Confidence Regions}
Under standard regularity assumptions, the nonlinear least-squares estimator is {\em asymptotically normal} \cite{vdVaart1998}. More specifically,
assume that {\newtext $\theta^* = [(\beta^*)\tr\!, (\gamma^*)\tr]\tr$} lies in the interior of the compact set $\Theta \subset \BR^{d_{\theta}}$, the regression function $f(\theta)\defeq \Phi(\beta)\gamma$ is twice continuously differentiable in a neighborhood of $\theta^*$, and that the noise sequence $\{W_t\}$ is i.i.d.\ with $\BE[W_t]=0$ and $\BE[W_t^2]=\sigma^2$. Assume further that the model is {\em identifiable}.
Then, any sequence of estimators $\hat{\theta}_n$ {\em globally minimizing} the {\em empirical} least‑squares objective over $\Theta$ is {\em consistent}, i.e., $\hat{\theta}_n \xrightarrow{p} \theta^*$ as $n \to \infty$. Finally, assuming that $R = \lim_{n\to\infty} R_n(\theta^*)$ is non-singular,
the
estimators satisfy
\begin{equation}
    \sqrt{n}(\hat{\theta}_n - \theta^*) \xrightarrow{d} \CN(0,\sigma^2R^{-1}).
\end{equation}
Consequently, as $n \xrightarrow{} \infty$, we have
\begin{align}
    \frac{n}{\sigma^2}(\hat{\theta}_n - \theta^*)\tr R (\hat{\theta}_n - \theta^*) \xrightarrow{d} \chi^2(d_{\theta}),
\end{align}
where $\chi^2(d_{\theta})$ is the chi-squared distribution with $d_{\theta}$ degrees of freedom. 
{\newtext For a prescribed confidence level $p$, let
$\mu_p \defeq F_{\chi^2(d_\theta)}^{-1}(p)$ denote the $p$-quantile of the
$\chi^2(d_\theta)$ distribution. Using the finite sample estimates of $R_n(\hat{\theta}_n) = \tfrac{1}{n}J(\hat{\theta}_n)\tr J(\hat{\theta}_n)$ and $\hat{\sigma}^2_n\defeq \tfrac{1}{n-d_{\theta}} \|y-\Phi(\hat{\beta}_n)\hat{\gamma}_n\|^2$,
{\newtext an asymptotic {\em confidence ellipsoid} can be constructed as
\begin{equation}
\label{equ:asym_conf_region}
\hspace*{-2mm}{\CA}_n^{(p)} \hspace{-0.6mm}\defeq \hspace{-0.6mm} \left\{ \hspace{-0.3mm}\theta
\in \BR^{d_{\theta}}\hspace{-0.7mm} :\hspace{-0.5mm} (\theta - \hat{\theta}_n)\tr\hspace{-0.5mm} R_n(\hat{\theta}_n) (\theta - \hat{\theta}_n) \hspace{-0.1mm}\leq\hspace{-0.1mm} \frac{\mu_p\hat{\sigma}_n^2}{n}
\right\}\!.\!\!
\end{equation}}
This construction guarantees
\(\lim_{n\to\infty}\mathbb{P}(\theta^* \in \CA_n^{(p)}) = p\).
}
\section{Non-Asymptotic Confidence Regions}
{\newtext In this section, we propose two non-asymptotic confidence region constructions for SNLLS problems. They are motivated by the {\em Sign-Perturbed Sums} (SPS) method \cite{Csaji2015} and its generalizations, the {\em Data Perturbation} (DP) methods \cite{KolumbanPhD2015}. DP is a family of hypothesis tests used to construct non-asymptotic confidence regions with {\em exact} coverage.
Although DP methods are related to {\newtext the} {\em wild bootstrap}, {\newtext in that} they typically resample residuals, they do not necessarily require {\em refitting} the model, yet they still guarantee {\em finite-sample coverage} (see \cite[Section V.D]{Csaji2015} for a comparison). Operating under very mild statistical assumptions, they require only a distributional invariance from the noise. The core idea is to generate alternative datasets via resampling, rank them using a 
performance measure, and reject the candidate model if the rank of the original dataset exceeds a given threshold.}

\begin{algorithm}[t]
    \vspace*{1mm}
	\caption{Pseudocode: {\newtext Initialization}\hspace{0.5mm}($p$)}
    \label{alg:sps_init}
	\begin{algorithmic}[1]
		\STATE Given the (rational) confidence probability $p \in (0,1)$, set integers $m > q >0$ such that $p = 1 - q/m$.
\STATE {\newtext Generate $G_1,\dots, G_{m-1} \sim \mathrm{Unif}(\mathcal{G})$ i.i.d.\ (uniform) random perturbations.
        See \eqref{equ:G_i_permutation} and \eqref{equ:G_i_sign_perturbation} for examples.}
		\STATE Generate (uniformly) a random permutation $\pi$ of the set $\{0,\dots, m - 1\}$, where each of the $m!$ possible permutations has probability $1/(m!)$ to be selected.
	\end{algorithmic}
\end{algorithm}
\begin{algorithm}[t]
	\caption{Pseudocode: RPG-Indicator\hspace{0.5mm}$(\theta)$}
    \label{alg:sps_indicator}
	\begin{algorithmic}[1]
		\STATE Compute the residual vector for $\theta$
        \vspace{-1mm}
        \begin{align*}
             r(\theta) \defeq y-\Phi(\beta)\gamma.
        \end{align*}
        \vspace{-5mm}
\STATE {\newtext Evaluate the reference and the perturbed functions:}
        \vspace{-1mm}
        {\newtext
        \begin{align*}
             &S_0(\theta)\, \defeq\, \tfrac{1}{n}
             J(\theta)\tr
             r(\theta),\\
             &S_i(\theta)\, \defeq\, \tfrac{1}{n}
             J(\theta)\tr
             G_i(r(\theta)),\quad i \in [\hspace{0.3mm}m-1\hspace{0.2mm}].
        \end{align*}}
        \vspace{-4mm}
		\STATE Compute the inverse {$R_n^{-1}(\theta)$} and the normalized rank 
of $\norm{S_0(\theta)}_{R_n^{-1}(\theta)}^2$ among {\newtext $\{\norm{S_i(\theta)}_{R_n^{-1}(\theta)}^2\}_{i=1}^{m-1}:$}
            \begin{align*}
                &\mathcal{R}(\theta)\, \defeq\, \\
                &\frac{1}{m}\Bigg[\,1+\sum_{i=1}^{m-1}\mathbb{I}\left(\norm{S_0(\theta)}_{R_n^{-1}(\theta)}^2 \succ_{\pi} \norm{S_i(\theta)}_{R_n^{-1}(\theta)}^2\right)\,\Bigg]\!,
            \end{align*}
        where ``$\succ_{\pi}$'' is ``$>$'' with random tie-breaking, more precisely, $\norm{S_k(\theta)}_{R_n^{-1}(\theta)}^2 \succ_{\pi} \norm{S_j(\theta)}_{R_n^{-1}(\theta)}^2$ if and only if $(\norm{S_k(\theta)}_{R_n^{-1}(\theta)}^2 > \norm{S_j(\theta)}_{R_n^{-1}(\theta)}^2) \lor (\norm{S_k(\theta)}_{R_n^{-1}(\theta)}^2 = \norm{S_j(\theta)}_{R_n^{-1}(\theta)}^2 \,\land\, \pi(k) > \pi(j))$.
        \vspace{2mm}
 		\STATE Return 1 if\, $\mathcal{R}(\theta) \leq p$, otherwise return 0.
	\end{algorithmic}
\end{algorithm}
\subsection{The {Residual-Perturbed Gradient} Algorithm}
Our first proposed DP method, called {\em Residual-Perturbed Gradient (RPG)}, consists of two parts: initialization, presented in Algorithm \ref{alg:sps_init}, and the indicator function, given in Algorithm \ref{alg:sps_indicator}. The initialization part computes the main parameters and the i.i.d.\ random perturbations given the desired confidence probability $p$ as input. 
The indicator function first computes the residual for the given candidate parameter vector, then forms a collection of perturbed residuals using the random perturbations. For each residual (original and perturbed) it computes a score vector using the Jacobian. Then, the original score vector is ranked among the randomized ones with performance measure $\|\cdot\|_{R_n^{-1}(\theta)}$, a scale-adjusted metric that defines the classical asymptotic confidence region, see \eqref{equ:asym_conf_region}. Using this indicator function, the $p$-level RPG confidence region can be defined as
\begin{equation}
        \CB_p\, \defeq\, \{\,\theta \in \BR^{d_{\theta}}\text{ : RPG-Indicator}(\theta) = 1\,\}.       
\end{equation}

Notice that, as a consequence of \eqref{equ:derivs_res_norm} and \eqref{equ:jacobian}, the quantity $\norm{S_0(\theta)}_{R_n^{-1}(\theta)}^2$  measures how strongly the first‑order optimality conditions are violated at $\theta$.
{\newtext In particular, at any {\em stationary point} of the least-squares objective \eqref{equ:snlls_matrix_form}, we have $S_0(\theta)=0$, and therefore such points are always in $\CB_p$.}

\subsection{The Global-Fit-Gap Algorithm}
\begin{algorithm}[t]
	\caption{Pseudocode: GFG-Indicator\hspace{0.5mm}$(\theta)$}
    \label{alg:gfgb_sps_indicator}
	\begin{algorithmic}[1]
        \vspace{1mm}
		\STATE Determine the residual vector for $\theta$,
        \vspace{-1mm}
        \begin{align*}
             r(\theta) \defeq y-\Phi(\beta)\gamma.
        \end{align*}
        \vspace{-5mm}
		\STATE Evaluate alternative outputs, for $i\in[m-1]$,
        \vspace{-1mm}
        \begin{align*}
             \tilde{y}_i(\theta)\, \defeq\, \Phi(\beta)\gamma+G_i(r(\theta)),
        \end{align*}
        and set $\tilde y_0(\theta)\defeq y$.
        \STATE Calculate global-refit estimates, for $i\in\{0,\dots,m-1\}$,
        \begin{equation*}
            (\hat{\beta}_i, \hat{\gamma}_i) \defeq \argmin_{\beta',\gamma'} \|\tilde{y}_i(\theta)-\Phi(\beta')\gamma'\|^2.
        \end{equation*}
        \vspace{-3mm}
        \STATE Determine global-fit-gap scores, for $i\in\{0,\dots,m-1\}$:
        \begin{align*}
            F_i(\theta) \defeq \|\tilde y_i(\theta)-\Phi(\beta)\gamma\|^2
            - \|\tilde y_i(\theta)-\Phi(\hat\beta_i)\hat\gamma_i\|^2.
        \end{align*}
        \vspace{-4mm}
\STATE Compute the normalized rank of $F_0(\theta)$ in $\{F_i(\theta)\}_{i=1}^{m-1}$,
        \begin{align*}
            \mathcal{R}(\theta)\, \defeq\, \frac{1}{m}\Bigg[\,1+\sum_{i=1}^{m-1}\mathbb{I}\!\left(F_0(\theta) \succ_{\pi} F_i(\theta)\right)\Bigg]\!,
        \end{align*}
        where ``$\succ_{\pi}$'' is ``$>$'' with random tie-breaking, as before.
\vspace{-2mm}
 		\STATE Return 1 if $\mathcal{R}(\theta) \leq p$, otherwise return 0.
	\end{algorithmic}
\end{algorithm}
While the RPG-Indicator is a natural extension of gradient-perturbation based DP methods
to SNLLS, it may be undesirable in highly {\em nonconvex} problems where several local minima 
can occur. To address this issue, we introduce an alternative construction, the {\em Global-Fit-Gap} (GFG), presented in Algorithm \ref{alg:gfgb_sps_indicator}, which evaluates candidate parameters based on {\em global optimality} rather than first-order conditions.

The GFG-Indicator generates alternative datasets by applying random perturbations to the residual, and for each dataset a nonlinear LS fit is performed. Then, it evaluates the input parameter by comparing its least-squares fit to that obtained after re-optimizing the model on the randomly perturbed residual data. {Hence, the resulting statistic quantifies the {\em suboptimality} of the input parameter for the corresponding dataset.} Using this algorithm, the GFG confidence region is 
\begin{equation}
        \confreg[p]\, \defeq\, \{\,\theta \in \BR^{d_{\theta}}\text{ : GFG-Indicator}(\theta) = 1\,\}.       
\end{equation}
{\newtext In the optimization step, any suitable nonlinear least-squares routine can be applied to obtain 
a global optimum.
Besides the found solution, no other stationary points will be automatically included in the resulting confidence set.
Compared to RPG, the GFG-Indicator requires repeated refitting of the model, and hence it is computationally more demanding, but it provides a stronger, more reliable acceptance criterion.}
\subsection{Illustration of RPG and GFG}
\begin{figure}[t]
  \vspace{1mm}
  \centering
  \includegraphics[scale=0.53]{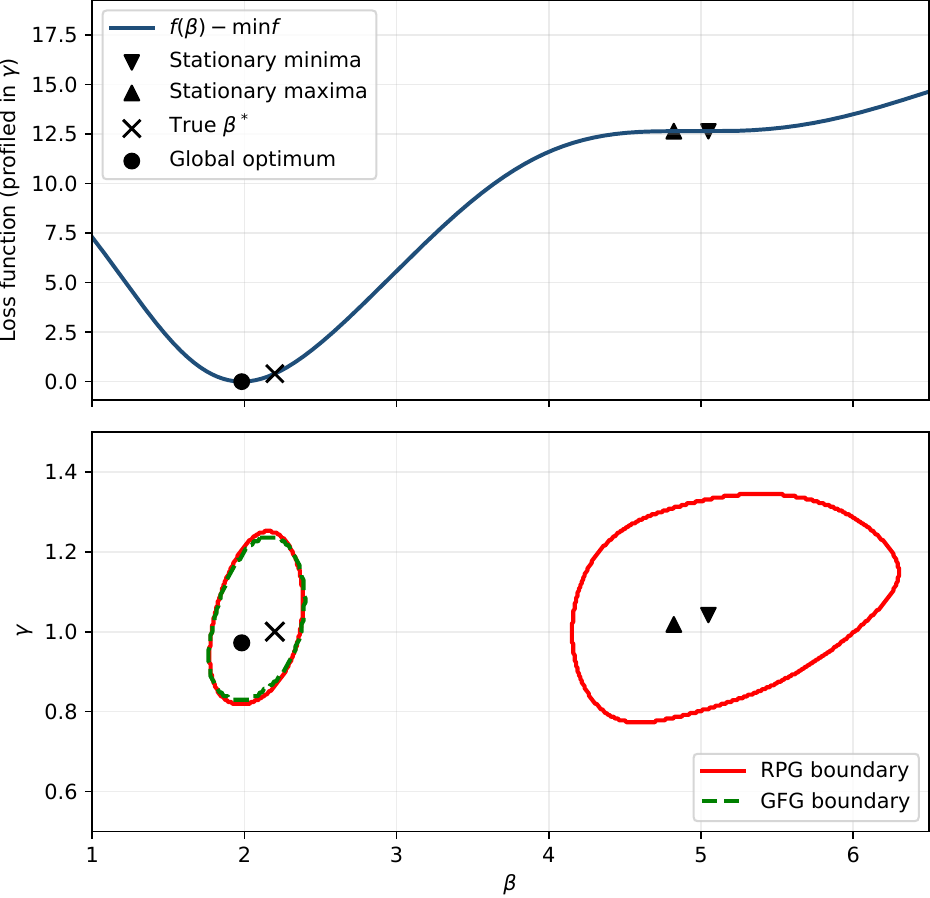}
\caption{{\newtext Comparison of the 0.5-level ($m=2$, $q=1$) RPG and GFG indicators for $n=72$, considering a $\mathrm{sinc}(x)$ function with Gaussian noise.}}
  \label{fig:sinc_example}
  \vspace{-2mm}
\end{figure}
We illustrate the difference between RPG and GFG using a scalar sinc function setup:
\begin{equation}
    Y_t = \gamma^* \frac{\sin(\beta^* X_t)}{\beta^* X_t} + W_t,
\end{equation}
where $\gamma^*=1$, $\beta^*=2.2$, $X_t$ is deterministic and equidistant on $[-2,2]$, $W_t \sim \CN(0, 0.55^2)$, and $n=72$. Since the noise is symmetric, we used sign-perturbations to construct the confidence regions. Fig.~\ref{fig:sinc_example} shows the profiled loss
in $\beta$, obtained after minimizing over $\gamma$, together with the $0.5$-level
($m=2$, $q=1$) acceptance boundaries of RPG and GFG.

{\newtext It can be observed that RPG and GFG constructed confidence regions that included the true parameter and the global optimum of the loss function, while RPG also constructed a separate region that included non-global stationary points.
In contrast, GFG did not provide additional acceptance regions. This illustrates that GFG can produce sharper confidence regions 
at the price of higher computational cost.}

\subsection{Theoretical Guarantees}
The constructed confidence regions $\CB_p$ and $\confreg[p]$ have exact coverage probability; therefore, the following theorem holds:
\begin{theorem}\label{thm:exact_confidence}
Under Assumptions \ref{assu:noise} and \ref{assu:full_rank_Jacobian}
    the coverage probability of the confidence regions $\CB_p$ and $\confreg[p]$ is exactly $p$:
\begin{equation}
            \BP(\theta^* \in \CB_p) = p, \quad \text{and} \quad \BP(\theta^* \in \confreg[p]) = p.
            \vspace{1mm}
    \end{equation}
\end{theorem}
\begin{proof}
    From \eqref{equ:derivs_res_norm}--\eqref{equ:res_derivs_at_true}, it follows that in the RPG and GFG indicator functions, for $\theta = \theta^* $:
    \begin{equation}
        S_i(\theta^*) = \tfrac{1}{n}
             J(\theta^*)\tr
             G_i(w),
    \end{equation}
    and
    \begin{equation}
    \tilde{y}_i(\theta^*) = \Phi(\beta^*)\gamma^*+G_i(w)
    \end{equation}
    for all $i \in \{0,\dots, m-1\}$, where 
$G_0$ is the identity element of $\CG$, i.e., the identity transformation.
{{\newtext Hence, for all $i \in \{0,\dots,m-1\}$, the quantities $S_i(\theta^*)$ and $\tilde{y}_i(\theta^*)$ depend on the perturbed noise sequences through the same functions, respectively,} which can be defined as $\psi_s(G_i(w)) \defeq S_i(\theta^*)$ and $\psi_y(G_i(w)) \defeq \tilde{y}_i(\theta^*)$. Then, the application of \cite[Theorem 2.17]{KolumbanPhD2015} guarantees that $\{\norm{S_i(\theta^*)}_{R_n^{-1}(\theta^*)}^2\}_{i=0}^{m-1}$ and $\{{F_i(\theta^*)}\}_{i=0}^{m-1}$ are {\em uniformly ordered}. In the theorem above, our two suggested methods have performance measures:
   \begin{equation}
       Z(\theta, x, y')=\|\tfrac{1}{n}
             J(\theta)\tr
             (y'-\Phi(\beta,x)\gamma)\|_{R_n^{-1}(\theta)}^2
   \end{equation} in the RPG case, and 
   \begin{equation}
       Z(\theta, x, y')=\|y'-\Phi(\beta,x)\gamma\|^2
            - \|y'-\Phi(\hat\beta,x)\hat\gamma\|^2
   \end{equation} in the GFG case, where we have written $x\defeq[X_1,\dots,X_n]\tr$ for the sake of completeness and $(\hat\beta,\hat\gamma) \defeq \argmin_{\beta',\gamma'} \|y'-\Phi(\beta',x)\gamma'\|^2$.
{\newtext Consequently, $\norm{S_0(\theta^*)}_{R_n^{-1}(\theta^*)}^2$ and $F_0(\theta^*)$ take each position in their respective ordering with the same probability $1/m$; therefore, their normalized rank is at most $p=1-q/m$ with probability exactly $p$.}}
\end{proof}

\section{Simulation Experiments}

{\renewcommand{\arraystretch}{1.4}
\newtext
\begin{table}[t]
\centering
\caption{Monte Carlo coverage in the GFIR model with \(95\%\) intervals.\vspace{-2mm}
}
\label{tab:m_sweep_coverage}
\scriptsize
\begin{tabular}{c c c c}
\hline
$m$ & $p=1-1/m$ & RPG empirical coverage & GFG empirical coverage\\
\hline
2  & 0.500 & 0.500\; [0.490, 0.510] & 0.497\; [0.487, 0.507] \\
5  & 0.800 & 0.799\; [0.791, 0.807] & 0.797\; [0.789, 0.805] \\
10 & 0.900 & 0.899\; [0.893, 0.905] & 0.896\; [0.890, 0.902] \\
20 & 0.950 & 0.946\; [0.942, 0.951] & 0.951\; [0.946, 0.955] \\
50 & 0.980 & 0.979\; [0.976, 0.982] & 0.982\; [0.979, 0.985] \\
\hline
\end{tabular}
\vspace{-2mm}
\end{table}}
In this section, we illustrate the RPG and GFG algorithms through two numerical experiments. In the first one we consider the identification of a {\em Generalized Finite Impulse Response}
(GFIR) system with {\newtext Laguerre basis functions} that can be used to model systems with {\em slowly decaying impulse responses}.
In the second experiment we consider a curve fitting problem with {\em Gaussian basis functions}{\newtext , a classical model in kernel methods and function approximation.}
\subsection{Generalized FIR Models with Adjustable Decay Rate}
We consider a 2-dimensional GFIR model using {\em Laguerre basis} functions with {\em adjustable decay rate}:
\begin{align}
Y_t = \gamma_1^{*}\varphi_1(\beta^{*},U_t) + \gamma_2^{*}\varphi_2(\beta^{*},U_t) + W_t,
\end{align}
where for $i \in \{1,2\}$: $\varphi_i(\beta,U_t) = L_i(z,\beta) U_t$, 
$z$ denotes the shift operator, and the filter is defined as
\vspace{-1mm}
\begin{align}
L_i(z,\beta)
=
\frac{1}{z-\beta}
\left(\frac{1-\beta z}{z-\beta}\right)^{\hspace{-0.5mm}i-1}\hspace{-4.5mm}.
\end{align}
In this experiment, the true parameters are $\beta^*=0.55$, $\gamma^*=[1.2, -0.8]\tr$, $\{W_t\}$ are i.i.d.\ heavy-tailed random variables sampled from a Pareto distribution with scale parameter $1$ and shape parameter $1.5$, the input $\{U_t\}$ is sampled from a uniform distribution with boundaries $[-2,2]$, and the sample size is $n=250$.
Notice that the resulting parameter vector is
$\theta^*=[\beta^*,(\gamma^*)\tr]\tr$, and the model is linear in $\gamma^*$ and nonlinear in $\beta^*$ through the Laguerre filters \(L_i(z,\beta)\), hence, the identification problem can be formulated as an SNLLS.

{\newtext For the construction of the confidence regions we used permutations, see \eqref{equ:G_i_permutation}.
The parameter estimate $(\hat\beta,\hat\gamma)$ was obtained by VP with Levenberg--Marquardt (LM) refinement.}

{\newtext To empirically check the finite-sample coverage claim and the dependence on $m$, we ran $10^4$ independent Monte Carlo trials, regenerating both the data and the randomized permutation transformations in each one. Table~\ref{tab:m_sweep_coverage} reports the fraction of trials for which $\theta^*\in\CB_p$ and $\theta^*\in\CC_p$ with exact $95\%$ Clopper--Pearson (CP) intervals. We used $q=1$, hence $p=1-q/m$, so increasing $m$ increases the nominal confidence level and gives more conservative rank regions. The empirical coverages closely follow the nominal levels for both methods. Moreover, in all cases, the nominal confidence level lies inside the corresponding exact $95\%$ CP interval.
}

\begin{figure}[t]
  \vspace{1mm}
  \centering
  \includegraphics[scale=0.45]{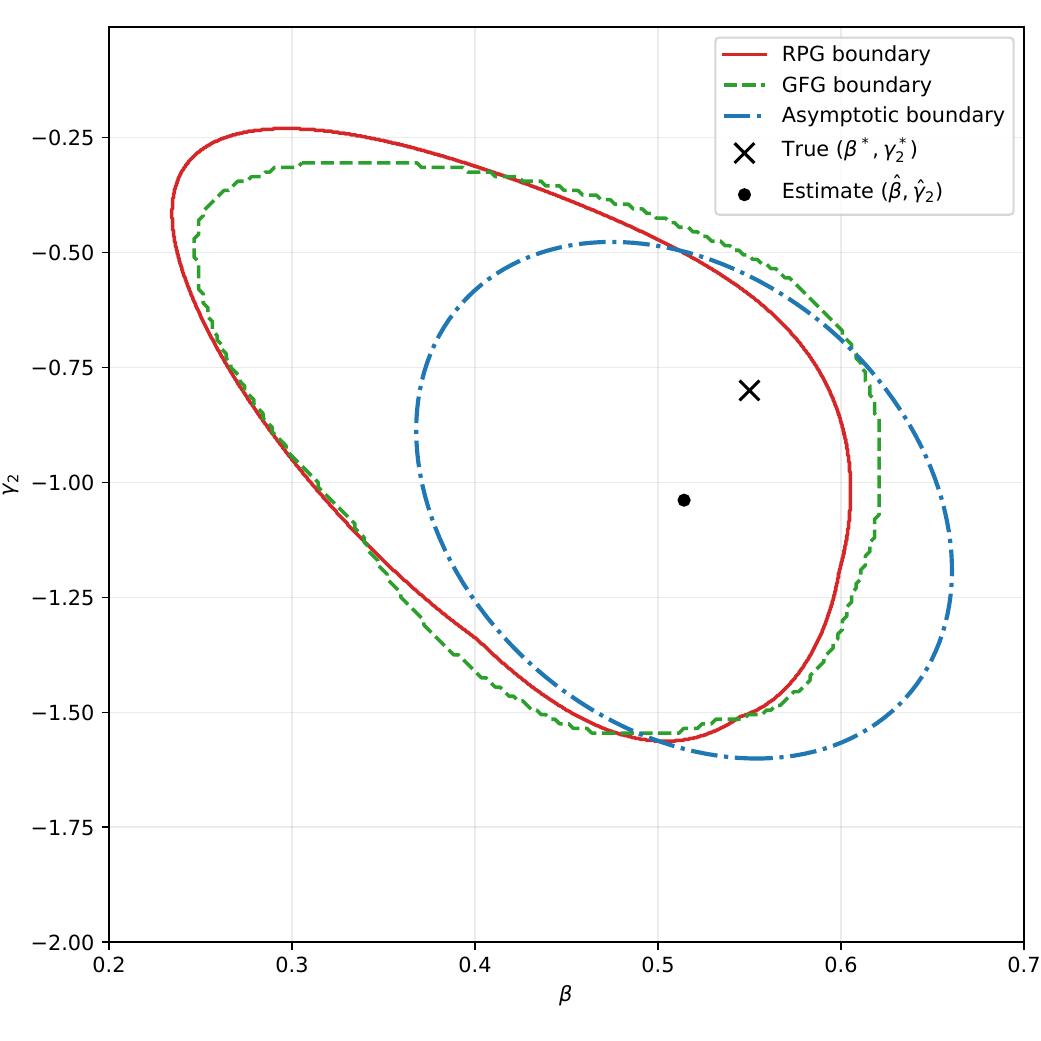}
  \vspace{-4mm}
  \caption{Comparison of 0.9-level RPG, GFG, and asymptotic confidence regions considering a GFIR model with heavy-tailed Pareto noise and fixed $\gamma_1=\hat\gamma_1$. $m=10$, $q=1$, $n=250$.}
  \label{fig:rtr_gfg_gfir}
  \vspace{-2mm}
\end{figure}

The $0.9$-level RPG and GFG confidence regions with $m=10$ and $q=1$, together with the asymptotic region, see \eqref{equ:asym_conf_region}, are illustrated in Fig.~\ref{fig:rtr_gfg_gfir}. In this figure we consider the $(\beta,\gamma_2)$ slice with $\gamma_1=\hat\gamma_1$. The contours of the indicator regions were obtained by evaluating them on a grid. 
It can be seen that the RPG and GFG regions have similar shapes and sizes, and both form a single connected component including the true parameter and the
estimate. Here, the RPG does not create additional disconnected
regions in the displayed slice. 
The ``asymptotic ellipsoid'' of \eqref{equ:asym_conf_region} is included only as a heuristic reference. It provides no finite-sample guarantees, and given that the noise terms have {\em infinite variance},
{\newtext it lacks asymptotic guarantees as well, since the classical central limit theorem does not apply to justify asymptotic normality}. 

\subsection{Gaussian Models with Adjustable Knots and Widths}
In this experiment, we consider the following separable nonlinear {\em Gaussian} model with {\em adjustable knots and widths}:
\begin{align}
Y_t = \sum_{k=1}^{d}\gamma_k^*\exp\left(
-\frac{\big(X_t-\mu_k^*\big)^2}{2(\sigma_k^*)^2}
\right) + W_t,
\end{align}
for $t\in[n]$, where $d=3$, the nonlinear parameter vector is
\begin{align}
    \beta^* &=
\begin{bmatrix}
\mu_1^* & \mu_2^* & \mu_3^* & \sigma_1^* & \sigma_2^* & \sigma_3^*
\end{bmatrix}\tr\notag\\
&=
\begin{bmatrix}
-2.0 & 0.45 & 2.0 & 0.42 & 0.70 & 0.55
\end{bmatrix}\tr,
\end{align}
and true linear parameters are
$\gamma^*=[1.8,-1.35,0.85]\tr$.
The input points $\{X_t\}$ are generated by uniform random sampling on $[-5,5]$,
the noises $\{W_t\}$ {\newtext are i.i.d.\ Laplace distributed} with zero mean and variance $1.5$, and 
$n=35$.

As the noises are symmetric about zero,
sign-perturbations are used, i.e., {\newtext $G_i = \text{diag}(\alpha_{i,1},\dots,\alpha_{i,n})$}, where for $i \in [\hspace{0.3mm}m-1\hspace{0.2mm}]$ and $t \in [\hspace{0.3mm}n\hspace{0.2mm}]$, $\BP(\alpha_{i,t} = 1) = \BP(\alpha_{i,t} = -1)\, =\, 1/2$.
Throughout this experiment we use $m=10$ and $q=1$.

\begin{figure}[t]
  \centering
  \vspace{1mm}
  \subfloat[RPG]{\includegraphics[width=0.47\textwidth]{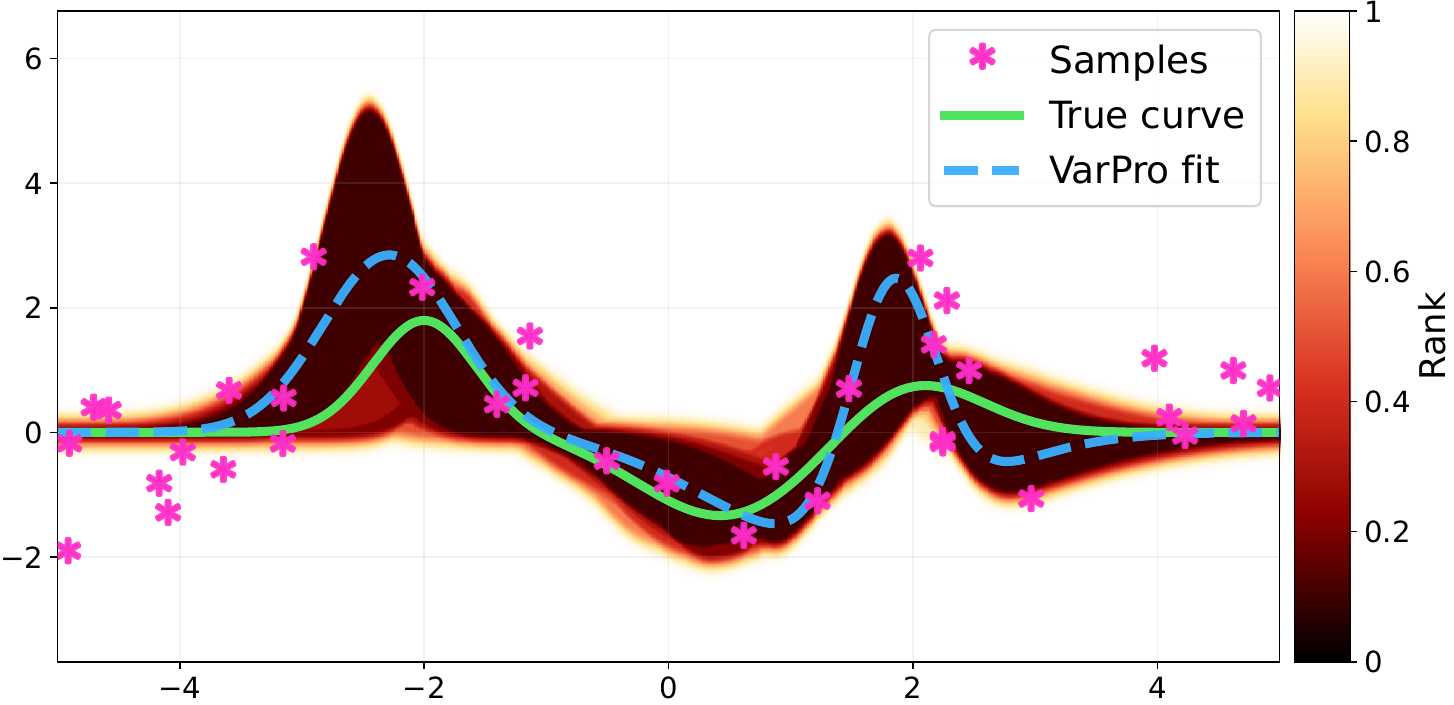}
    \label{fig:rpg_gfg_cf_a}
  }
  \vspace{-1mm}
  \subfloat[GFG]{\includegraphics[width=0.47\textwidth]{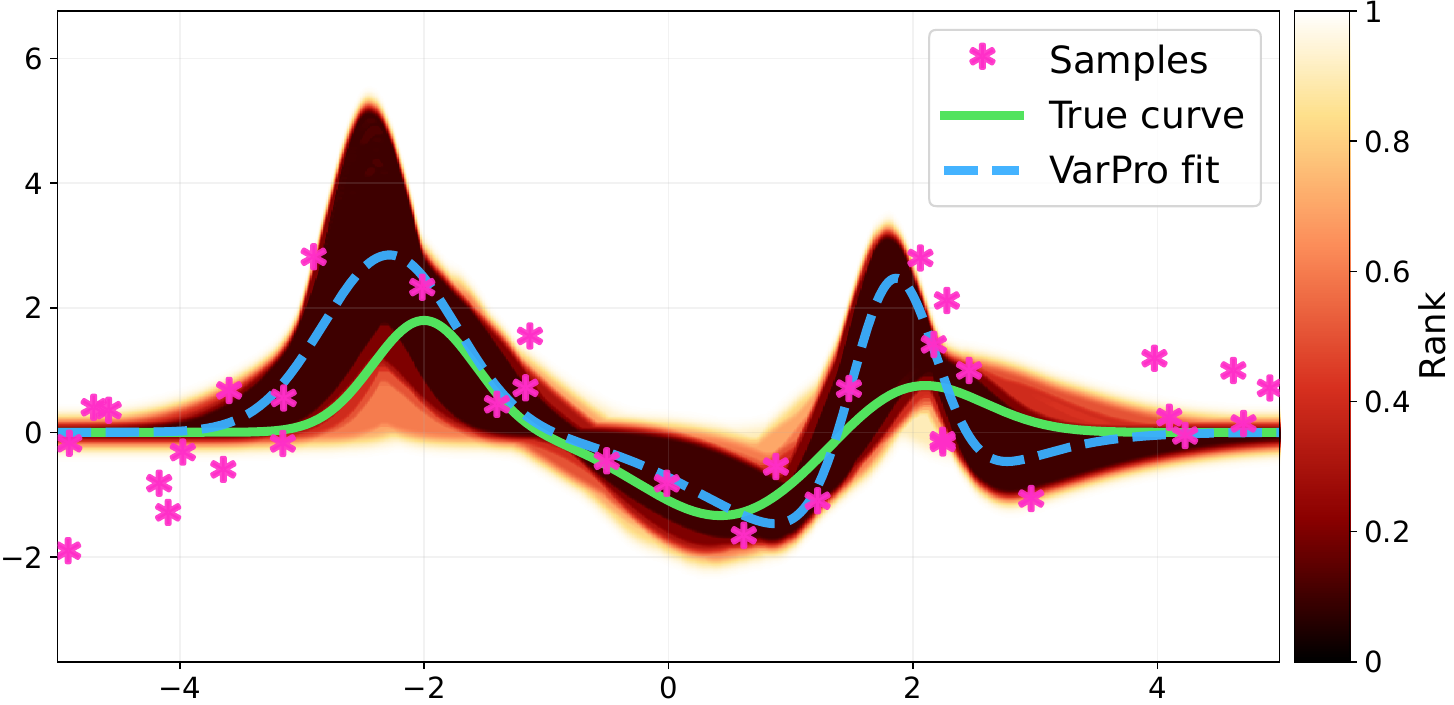}
    \label{fig:rpg_gfg_cf_b}
  }
  \caption{Comparison of RPG and GFG rank maps considering a Gaussian model with
adjustable knots, widths, and {\newtext Laplace} noise. $n=35$, $m=10$. For visual clarity, the displayed rank maps use the same post-processing.}
  \label{fig:rpg_gfg_cf}
  \vspace{-3mm}
\end{figure}

Fig.~\ref{fig:rpg_gfg_cf} shows the noisy samples together with the true and estimated curve, obtained by VP, as well as the corresponding rank maps of RPG and GFG. To compute these rank maps, we evaluated both methods on a structured set of candidate
nonlinear parameter vectors constructed around the true and VP-estimated parameters by interpolating between them. For each candidate, the linear coefficients are recomputed by LS and the resulting
curve is mapped into the input-output plane.

{\newtext It can be observed that the GFG rank map is locally more concentrated around the fitted curve. This can be seen near $x\approx -2.2$ and on the right shoulder around $x\approx 3$, where RPG spreads confidence over a broader family of curves. Hence, GFG yields a sharper local characterization of uncertainty,
at the cost of repeated refits.
}

\section{Conclusions}
{\newtext 
SNLLS problems are widely used in control theory and machine learning, with the VP method as their archetypal solution. Despite their popularity, there are very few results on uncertainty quantification (UQ) for SNLLS. 
In this paper, we introduced two new methods, RPG and GFG, for constructing distribution-free confidence regions in SNLLS problems. By leveraging the group invariance of the noise, such as symmetry or exchangeability, we have shown that it is possible to move beyond traditional asymptotic ellipsoids, which often provide unreliable guarantees in finite-sample regimes.
We proved that both methods guarantee non-asymptotically exact coverage,
requiring only mild assumptions on the noise distribution. Numerical experiments further highlighted the practical trade-offs between the two approaches: while RPG is more computationally efficient, it may capture local extrema in highly non-convex landscapes. GFG provides more selective confidence regions by focusing on global optimality, albeit at a higher computational cost.}

\section*{Acknowledgments}
The authors would like to thank S\'andor Kolumb\'an for the valuable discussions on data perturbation methods.

\bibliographystyle{IEEEtran}
\bibliography{mybib}

@article{bertozzi2024asymptotic,
  title={Asymptotic behaviour of the confidence region in orbit determination for hyperbolic maps with a parameter},
  author={Bertozzi, Nicola and Bonanno, Claudio},
  journal={Physica D: Nonlinear Phenomena},
  volume={470},
  pages={134403},
  year={2024},
  publisher={Elsevier}
}

@article{kundu1999asymptotic,
  title={On asymptotic behavior of least squares estimators and the confidence intervals of the superimposed exponential signals},
  author={Kundu, Debasis and Mitra, Amit},
  journal={Signal Processing},
  volume={72},
  number={2},
  pages={129--139},
  year={1999},
  publisher={Elsevier}
}

@PHDTHESIS{KolumbanPhD2015,
  author = {S\'andor Kolumb\'an},
  title = {System identification in highly non-informative environment},
  school = {Budapest University of Technology and Economics, Hungary, and Vrije Universiteit Brussels, Belgium},
  year = {2015}
}

@article{Csaji2015,
  title={\relax{Sign-Perturbed Sums}: A New System Identification Approach for Constructing Exact Non-Asymptotic Confidence Regions in Linear Regression Models},
  author={B. {\relax Cs}. Cs\'aji and M.~C. Campi and E. Weyer},
  journal={IEEE Transactions on Signal Processing},
  volume={63},
  number={1},
  pages={169--181},
  year={2015},
  publisher={IEEE}
}

@BOOK{vdVaart1998,
  author = {A. W. van der Vaart},
  title = {Asymptotic Statistics},
  publisher = {Cambridge University Press},
  year = 1998}

@article{Golub1973,
author = {Golub, G. H. and Pereyra, V.},
title = {The Differentiation of Pseudo-Inverses and Nonlinear Least Squares Problems Whose Variables Separate},
journal = {SIAM Journal on Numerical Analysis},
volume = {10},
number = {2},
pages = {413-432},
year = {1973},
}

@article{golub2003,
  title={Separable nonlinear least squares: the variable projection method and its applications},
  author={Golub, Gene and Pereyra, Victor},
  journal={Inverse problems},
  volume={19},
  number={2},
  pages={R1},
  year={2003},
  publisher={IOP Publishing},
}

@ARTICLE{VP_machine_learning,
  author={Gan, Min and Guan, Yu and Chen, Guang-Yong and Chen, C. L. Philip},
  journal={IEEE Transactions on Neural Networks and Learning Systems}, 
  title={Recursive Variable Projection Algorithm for a Class of Separable Nonlinear Models}, 
  year={2021},
  volume={32},
  number={11},
  pages={4971-4982},
  }

@ARTICLE{SNLLapp1,
  author={Chen, Guang-Yong and Gan, Min and Wang, Shuqiang and Chen, C. L. Philip},
  journal={IEEE Transactions on Image Processing}, 
  title={Insights Into Algorithms for Separable Nonlinear Least Squares Problems}, 
  year={2021},
  volume={30},
  number={},
  pages={1207-1218},
}

@ARTICLE{VP_L1,
  author={Chen, Guang-Yong and Su, Xiang-Xiang and Gan, Min and Guo, Wenzhong and Chen, C. L. Philip},
  journal={IEEE Transactions on Automatic Control}, 
  title={Robust Variable Projection Algorithm for the Identification of Separable Nonlinear Models}, 
  year={2024},
  volume={69},
  number={9},
  pages={6293-6300},
}

@article{VPlargescale,
title = {Variable Projection algorithms: Theoretical insights and a novel approach for problems with large residual},
journal = {Automatica},
volume = {177},
pages = {112300},
year = {2025},
issn = {0005-1098},
author = {Guangyong Chen and Peng Xue and Min Gan and Jing Chen and Wenzhong Guo and C.L. Philip Chen},
}

@ARTICLE{VP_RBF-ARX,
  author={Gan, Min and Li, Han-Xiong and Peng, Hui},
  journal={IEEE Transactions on Cybernetics}, 
  title={A Variable Projection Approach for Efficient Estimation of {RBF-ARX} Model}, 
  year={2015},
  volume={45},
  number={3},
  pages={462-471},
}

@article{nonsmoothVP,
author = {van Leeuwen, Tristan and Aravkin, Aleksandr Y.},
title = {Variable Projection for NonSmooth Problems},
journal = {SIAM Journal on Scientific Computing},
volume = {43},
number = {5},
pages = {S249-S268},
year = {2021},
}

@article{VPtrain,
author = {Newman, Elizabeth and Ruthotto, Lars and Hart, Joseph and van Bloemen Waanders, Bart},
title = {Train Like a {(Var)Pro}: Efficient Training of Neural Networks with Variable Projection},
journal = {SIAM Journal on Mathematics of Data Science},
volume = {3},
number = {4},
pages = {1041-1066},
year = {2021},
}

@ARTICLE{WVPNet,
  author={Dózsa, Tamás and Böck, Carl and Meier, Jens and Kovács, Péter},
  journal={IEEE Transactions on Neural Networks and Learning Systems}, 
  title={Weighted {H}ermite Variable Projection Networks for Classifying Visually Evoked Potentials}, 
  year={2025},
  volume={36},
  number={7},
  pages={12415-12428},
}

@inproceedings{VPident,
author = {Tamás Dózsa and Mátyás Szabari and Alexandros Soumelidis and Péter Kovács},
title = {Pole Identification using discrete Laguerre expansion and variable projection},
volume = {56},
number = {2},
pages = {3994-3999},
year = {2023},
booktitle = {22nd IFAC World Congress},
}

\end{document}